# Detailed Primary and Secondary Distribution System Model Enhancement Using AMI Data

Karen Montano-Martinez, *Student Member, IEEE,* Sushrut Thakar, *Student Member, IEEE,* Shanshan Ma, *Member, IEEE,* Zahra Soltani, *Student Member, IEEE,* Vijay Vittal, *Life Fellow, IEEE*, Mojdeh Khorsand, *Member, IEEE,* Raja Ayyanar, *Senior Member, IEEE,* and Cynthia Rojas, *Member, IEEE*

*Abstract*— Reliable and accurate distribution system modeling, including the secondary network, is essential in examining distribution system performance with high penetration of distributed energy resources (DERs). This paper presents a highly automated, novel method to enhance the accuracy of utility distribution feeder models to capture their performance by matching simulation results with corresponding field measurements. The method is demonstrated using an actual feeder from an electrical utility with high penetration of DERs. The method proposed uses advanced metering infrastructure (AMI) voltage and derived active power measurements at the customer level, and data acquisition systems (DAS) measurements at the feeder-head, in conjunction with an AC optimal power flow (ACOPF) to estimate customer active and reactive power consumption over a time horizon, while accounting for unmetered loads. Additionally, the method proposed estimates both voltage magnitude and angle for each phase at the unbalanced distribution substation. Furthermore, the accuracy of the method developed is verified by comparing the time-series power flow results obtained from the enhancement algorithm with OpenDSS results. The proposed approach seamlessly manages the data available from the optimization procedure through the final model verification automatically.

## Nomenclature

### A. Sets and Indices

| | |
|---|---|
| $\Omega_N$ | Set of buses. |
| $\Omega_D$ | Subset of buses with load. |
| $\Omega_{AMI_V}$ | Subset of load buses with AMI voltage information. |
| $\Omega_{AMI_D}$ | Subset of load buses with AMI active power information. |
| $\Omega_{D1}$ | Subset of load buses without AMI information ($\Omega_D - \Omega_{AMI_D}$). |
| $\Omega_{PV}$ | Subset of buses with PV resources. |
| $\Omega_{cap}$ | Subset of buses with capacitor banks. |
| $\Omega_{SB}$ | Set of substation buses. |
| $\Omega_L$ | Set of distribution lines. |
| $\Omega_H$ | Set of feeder-head buses. |
| $\Omega_T$ | Set of distribution transformers. |
| $\psi$ | Set of phases, i.e., $\{a, b, c\}$. |
| $\delta(i)$ | Set of bus nodes connected to bus $i$. |

### B. Parameters

| | |
|---|---|
| $B_{c,\phi}$ | Capacitance of capacitor bank $c$ on phase $\phi$. |
| $d_l^D$ | Gross load at bus $l$ with AMI active power measurements. |
| $P_l^D$ | AMI active power delivered to bus $l$. |
| $P_l^R$ | AMI active power received at bus $l$. |
| $P_l^{PV}$ | AMI active power produced by the solar PV at bus $l$. |
| $P_m^{Tr}$ | No-load loss of transformer $m$. |
| $P^H/Q^H$ | Minimum/maximum power factor for load $l$. |
| $|\hat{V}_{i,\phi}|$ | DAS active/reactive power data at feeder-head. |
| $R_{i,j}^{\phi,p}$ | AMI voltage magnitude data at phase $\phi$ at bus $i$. |
| $X_{i,j}^{\phi,p}$ | Resistance of distribution line $(i,j)$ between phases $\phi$ and $p$. |
| $y_{i,j}^{\phi,p}$ | Reactance of distribution line $(i,j)$ between phases $\phi$ and $p$. |
| | Admittance of distribution line $(i,j)$ between phases $\phi$ and $p$. |

### C. Variables

| | |
|---|---|
| $|V_{s,\phi}|/\angle V_{s,\phi}$ | Magnitude/angle of voltage at phase $\phi$ at substation $s$. |
| $V_{i,\phi}^r/V_{i,\phi}^{im}$ | Real/imaginary part of voltage at phase $\phi$ at bus $i$. |
| $I_{i,\phi}^{r,inj}/I_{i,\phi}^{im,inj}$ | Real/imaginary part of injected current at phase $\phi$ at bus $i$. |
| $I_{i,j,\phi}^r/I_{i,j,\phi}^{im}$ | Real/imaginary part of line flow current at phase $\phi$ in line $(i,j)$. |
| $P_{s,\phi}^G/Q_{s,\phi}^G$ | Active/reactive power output at phase $\phi$ at substation $s$. |
| $P_{l,\phi}^D/Q_{l,\phi}^D$ | Active/reactive power at phase $\phi$ for load $l$. |
| $Q_{c,\phi}^C$ | Reactive power output of capacitor bank $c$ at phase $\phi$. |

Manuscript submitted May 25, 2021. This work was supported in part by Advanced Research Projects Agency-Energy under award number DE-AR00001858-1631 and by the U.S. Dept. of Energy, Solar Energy Technologies Office, under award number. DE-EE0008773.

K. Montano-Martinez, S. Thakar, S. Ma, Z. Soltani, V. Vittal, M. Khorsand, and R. Ayyanar are with the School of Electrical, Computer and Energy Engineering, Arizona State University, Tempe, AZ 85287 USA (e-mail: kvmontan@asu.edu;　　　sushrut.thakar@asu.edu;　　　shansh19@asu.edu; zsoltani@asu.edu;　　　vijay.vittal@asu.edu;　　mojdeh.khorsand@asu.edu; rayyanar@asu.edu).

C. Rojas is with Arizona Public Service, Phoenix, AZ 85072 USA (e-mail: cynthia.rojas@aps.com).

## I. INTRODUCTION

THE increased penetration of distributed energy resources (DERs) – which include renewable energy resources, distributed energy storage, and electric vehicles (EVs) – in the electric grid has resulted in unprecedented changes to power system operation, such as bidirectional power flows and increased voltage fluctuations [1]. As DERs continue to grow, these issues would further impact the power distribution system's planning and operation, increasing the need for monitoring and controlling these resources [2].

Conventionally, distribution systems with just one source at the substation have relied on major model approximations, avoiding detail extending to the secondary circuits [3], [4]. However, a large share of DERs are located at the distribution system secondary at on-site customer locations, creating the need for a paradigm shift to model distribution systems with more accuracy [5]. Several authors have proposed methods to create approximate models of the secondary circuit [6]–[9]. Nevertheless, an inaccurate model of the secondary network can misrepresent the effects of DERs resulting in different voltages and incorrect power calculation across the secondary network when conducting power flow analysis [10]. Therefore, reliable and accurate distribution system modeling, including the secondary network and the various components such as load and DER are essential for distribution system operational analysis while accommodating a high level of DERs.

Due to the recent emphasis on a more accurate representation of the grid, many utilities now have extensive geographic information system (GIS) databases on feeder equipment and conductor segments. Additionally, as technology is transforming, advanced metering infrastructure (AMI) and data acquisition systems (DAS) are expanding on the distribution network. By leveraging these data, a high-fidelity feeder model can be developed to address the needs of the utilities to improve distribution system modeling to effectively plan and operate for future smart distribution systems with DERs.

This increase in distribution system visibility due to AMI and other emerging sensors has raised the interest in new methods to accurately model the distribution network [11]. A distribution system parameter estimation (DSPE) method using optimal linear regression model and AMI data is proposed in [12]. Nevertheless, the authors validated their method using test cases and assumed the secondary network to be radial, balanced, and assumed the availability of reactive power measurements. In reality, none of these assumptions are always true; therefore, this method portrays an approximate representation of the secondary. Similarly, the authors in [13], [14] use GIS and AMI data to model single-phase loads and high penetration of DERs. However, the authors only modeled loads where measurements exist and assumed a constant power factor for all loads.

A method to estimate the impedance of secondary branches using AMI measurements of voltage, and active and reactive power is proposed in [15]. The authors presented an optimization algorithm based on gradient search to calculate the voltage of the upstream node from a measured load. However, this approach requires complete visibility of all the loads in the feeder to create an accurate model and may be inaccurate for the feeders with unmetered loads.

This paper focuses on the estimation and modeling of distribution feeder's loads (active and reactive components), customer voltages, DERs output, and substation voltages, active, and reactive power. This paper presents an AC optimal power flow (ACOPF) formulation. The ACOPF uses a formulation based on a three-phase current-voltage (IV) model for unbalanced distribution networks with mutual impedance modeling, originally developed in [16]. The optimization-based modeling method of this paper is novel and can be extended without loss of generality to model any distribution system.

The key contributions of this paper are multifold:

1) This paper presents a novel computationally efficient method for estimating customer active and reactive power time-series consumption using AMI voltage and derived power measurements at some customer locations along the feeder. The method also estimates unmetered load parameters by knowing their GIS location.

2) This method provides a complete power flow solution, including the secondary circuit representation, using sparse measurements along a feeder, extending the observability and planning capabilities of the feeder under study.

3) An unbalanced substation model is proposed and implemented to capture the unbalanced nature of distribution substations in practical utility feeders.

4) The method proposed uses DAS measurements at the head of the feeder to estimate both voltage magnitude and angle at the substation for each phase. The active power and reactive power flow at the substation are also determined for each phase.

5) This method is demonstrated on an actual three-phase feeder with high penetration of DERs from an electric utility. The validation conducted closely matches the field measurements (AMI and DAS measurements) in terms of quantities at the substation and in terms of the voltages along the feeder at the individual residences.

6) The method developed is shown to be accurate when comparing the power flow obtained from the optimization-based algorithm against the power flow from the open-source software OpenDSS (Open Distribution System Simulator) [17]. Therefore, the method developed can be used to enhance any OpenDSS feeder model.

7) The objective function used in the proposed method can be modified and extended according to a specific necessity while obtaining an accurate power flow.

The proposed approach seamlessly transfers the data available through the optimization-based method to the final model verification, with limited human intervention. The method is also computationally efficient since no iterative power flow solutions are carried out. This paper serves as a benchmark for further improving the model accuracy of distribution systems for analysis and operation of distribution system networks with high DERs penetration at the utility level.

This paper is structured as follows. Section II describes the modeling data resources available at the utility level and its implementation in the development of a detailed distribution system model. Section III presents the proposed distribution system model enhancement algorithm, the data implementation to create the input data management for the algorithm, and the validation of the power flow results for the algorithm enhanced model against OpenDSS. Section IV and Section V demonstrate the performance of the proposed method on a utility feeder with high solar photovoltaic (PV) penetration and Section VI concludes the paper.

## II. MODELING DATA RESOURCES

The objective of the proposed model enhancement method is to obtain an accurately detailed distribution system model to capture the performance of the feeder by matching simulation results with corresponding field measurements. Fig. 1 illustrates the modeling data resources for the proposed distribution system model enhancement method. The GIS database contains conductor details and the latitude and the longitude of both endpoints of all the circuit sections, including the secondary network. Additionally, equipment ratings of transformers and capacitors, and locations of the system elements such as loads, and PV units are available. These data are used to create the distribution system topology, including load allocation, using the approach presented in [18], [19]. The DAS database contains hourly feeder-head measurements, which are used for construction and validation of the distribution feeder model.

The AMI database contains measurements of energy $(kWh)$ from 784 PV meters (production meters) and 1652 load meters (billing meters). These measurements are given as aggregated values each 15 minutes or each hour depending on the meter. The active power $(kW)$ is derived on each meter by aggregating the measurements of energy consumption for one hour. The AMI database also includes voltage magnitude measurements from 1194 load meters given each 15 minutes. The voltage measurements corresponding to every hour are considered for validation purposes. These considerations are taken after discussing with the utility who provided the measurements about how they manage their data. There are two cases for load definition: metered loads $(\Omega_{AMI_D})$ and unmetered loads $(\Omega_{D1})$. For the metered loads, the active power definition is based on the AMI measurements of the derived active power available at some premises. Fig. 2 shows the metering infrastructure installed at a typical metered premise, where the household has separate meters for PV production and billing. The load active power definition for metered loads is represented by the gross load $(d_l^D, l \in \Omega_{AMI_D})$, which is the total active power demand at a household. The production meter measurements are used to derive the active power produced by the PVs $(P_l^{PV}, l \in \Omega_{PV})$ and to define the PV generation. The billing meter is a bi-directional meter whose measurements are used to derive the power delivered by the utility to the customer $(P_l^D)$ and the power received by the utility from the customer $(P_l^R)$. Hence, (1) is used to estimate the gross load of a metered premise.

$$d_{l,\phi}^D = P_{l,\phi}^D - P_{l,\phi}^R + P_{l,\phi}^{PV}, \forall \ l \in \Omega_{AMI_D}, \phi \in \psi(l) \quad (1)$$

On the other hand, the power definition for unmetered loads $(P_l^D, l \in \Omega_{D1})$ is set as variable to be estimated by the enhancement method.

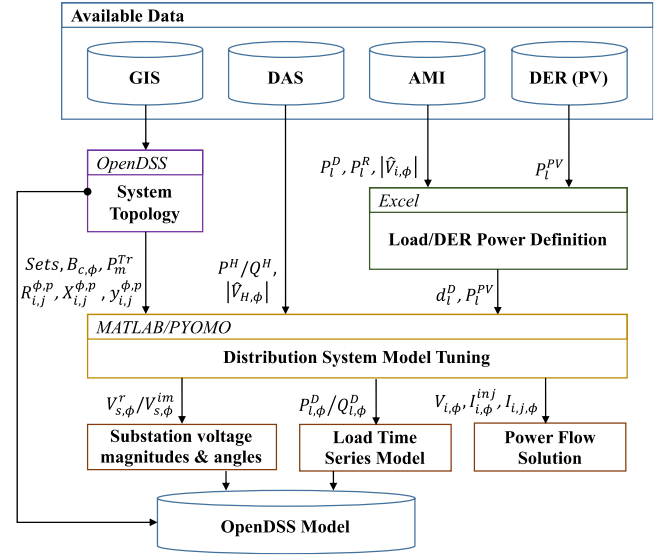

Fig. 1. Modeling data resources for distribution system model enhancement.

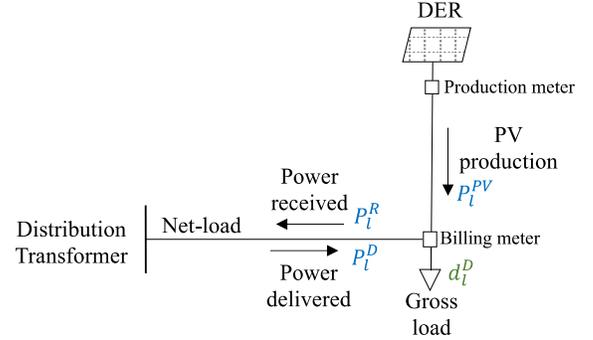

Fig. 2. Metering infrastructure at a typical premise: available measurements are in blue, estimated values are in green.

The proposed method relies on the following assumptions regarding the available data:

1) The secondary network topology is assumed to be known. If the topology is unknown, the approach presented in [18] can be used to estimate the system topology.

2) Feeder-head measurements are assumed to be available. These measurements may include but not be limited to feeder-head total active power $(P^H)$ and reactive power $(Q^H)$ measured on a three-phase basis, and feeder-head voltage magnitude $(|\hat{V}_{i,\phi}|, i \in \Omega_H)$ for at least one phase, $\phi$.

3) The gross load of the metered loads $(d_l^D, l \in \Omega_{AMI_D})$ is assumed to be available or derivable from the AMI measurements available in the system.

4) The active power of the DERs $(P_l^{PV}, l \in \Omega_{AMI})$ is assumed to be available from the AMI measurements available in the system.

5) Load voltage measurements $(|\hat{V}_{i,\phi}|, \forall i \in \Omega_{AMI_V})$ are assumed to be available for validation purposes. However, it is not necessary to have these voltage measurements for the same loads as in Assumption 3.

The system topology, feeder-head measurements, load voltage measurements available, DERs power definition, and the initial power definition of the loads are processed in MATLAB to create the input data to the optimization-based method. The ACOPF based on an IV formulation for distribution system model enhancement is programmed using Pyomo which is a Python-based, open-source optimization modeling language [20], [21]. IPOPT is used as the non-linear solver for the proposed ACOPF approach [22]. The output from the distribution system model enhancement includes time-series substation voltage magnitudes and angles, load model, and power flow solution. To validate the accuracy of the proposed optimization-based method, the power flow from the resulting enhanced distribution system model is compared with the power flow from OpenDSS.

## III. DISTRIBUTION SYSTEM MODEL ENHANCEMENT ALGORITHM

This section presents the formulation of the proposed method, the data implementation to create the input data for the algorithm, the formulation of the proposed ACOPF approach, the algorithm output and simulation capabilities, and the power flow validation of the resulting enhanced model against OpenDSS. A flowchart explaining the optimization technique proposed is summarized in Fig. 3.

### A. Input Data and Initialization

The input data contains the feeder topology information, and AMI/DAS measurement data. The feeder topology data includes the information to create the subset of buses with load, PV, capacitor and transformers, and impedance information of each distribution line. The AMI measurement data includes the energy measurements ($kWh$) used to derive the hourly active power ($kW$) for the metered load. The AMI database also includes voltage magnitude measurements for some load in the feeder. The DAS measurement data includes total active/reactive power at the feeder-head, and the voltage magnitude of one phase. The detailed procedure for creating the input data is described in Algorithm 1.

Most nonlinear solvers only find local optimal solutions for nonconvex problems [23]. Therefore, an adequate initialization is essential to find a solution that meets the problem requirements. After the input data is created, the parameters and variables are defined. The parameters are listed under the assumptions in Section II and are kept constant during all the simulations. The variables in the enhancement algorithm include bus voltages, unmetered load active power definition, load reactive power definition, bus injection currents, line flow currents, capacitors reactive power production, and per phase substation active/reactive power and voltage. The feeder modeled in this paper corresponds to a purely residential Arizona utility feeder, therefore the variables for the proposed distribution system model enhancement method are initialized accordingly. Nevertheless, this initialization can be adjusted without loss of generality knowing whether the load is residential, commercial, or industrial. The bus voltages, bus injection currents, and line flow currents are initialized using a flat start. A typical household peak load in Arizona stands between $4 - 7\ kW$ due to the need for air conditioning, therefore a value of $5\ kW$ is chosen as an initial value for the

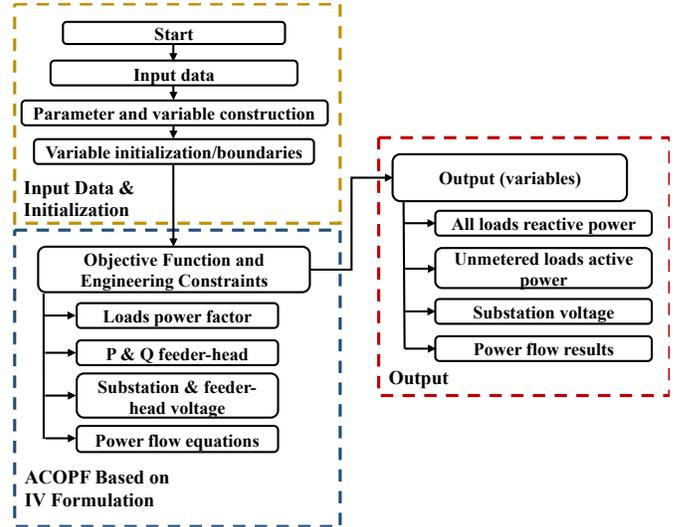

Fig. 3. Flowchart for distribution system model enhancement algorithm.

---

**Algorithm 1:** Input data creation

**Input:** Feeder topology and AMI/DAS measurements;

**Output:** Distribution system model enhancement algorithm input data;

Read feeder topology and line impedances ($R_{i,j}^{\phi,p}, X_{i,j}^{\phi,p}, y_{i,j}^{\phi,p}$);

Create topology sets ($\Omega_N, \Omega_D, \Omega_{PV}, \Omega_{cap}, \Omega_{SB}, \Omega_L, \Omega_H, \Omega_T$);

Read equipment information ($B_{c,\phi}, P_m^{Tr}$);

Read DAS measurements ($P^H, Q^H, |\hat{V}_{H,\phi}|$);

Read AMI measurements of energy ($kWh$) and voltage magnitude ($|\hat{V}_{i,\phi}|$);

Create sets of buses with measurements ($\Omega_{AMV}, \Omega_{AMI_D}$);

**FOR** hour $h$ in 1…24 **DO**

  Derive hourly active power measurements at each meter by aggregating the measurements of energy consumption ($P_l^D, P_l^R, P_l^{PV}, \forall\ l \in \Omega_{AMI_D}$);

  Derive the gross load ($d_l^p, l \in \Omega_{AMI_D}$) using (1);

  Read the PV active power ($P_l^{PV}, l \in \Omega_{PV}$);

  Write gross load and PV active power data;

  Write line/transformer data ($R_{i,j}^{\phi,p}, X_{i,j}^{\phi,p}, y_{i,j}^{\phi,p}$) for the algorithm;

  Write equipment data ($B_{c,\phi}, P_m^{Tr}$) for the algorithm;

  Write voltage measurement data ($|\hat{V}_{i,\phi}|$) for the algorithm;

**END FOR**

---

active power of unmetered loads. The reactive power for all the loads in the feeder is initialized using a power factor of 0.9. The capacitor reactive power is initialized using its nominal capacitance. The substation per phase active/reactive power are initialized using one-third of the total three-phase measurement at the feeder-head. The substation voltage is initialized at the same voltage as the feeder-head measurement.

### B. Enhancement Algorithm: ACOPF Based on Current-Voltage (IV) Formulation

This subsection formulates the optimization-based problem proposed for the distribution system model enhancement method as an ACOPF formulation based on a three-phase IV model for unbalanced distribution networks with mutual impedance which is more appropriate for distribution networks [16], [24]. The IV formulation solves a linear system of equations without decomposition, unnecessary constraints or omissions, and it may be computationally easier to solve than the traditional quadratic power flow formulations [23]. The ACOPF presented in this paper, co-optimizes active and reactive power along the distribution feeder. The nonlinear formulation is carried out in rectangular coordinates. The model enhancement algorithm proposed reads the input data and initialization of variables and transfers the information through the engineering constraints while optimizing the system performance according to the objective function, shown in Fig. 3.

The objective of this formulation is to minimize the norm of the difference between bus voltage magnitudes and the corresponding AMI voltage measurement data. Since IV formulation of power flow is considered, the square of the voltage magnitude is used, which is equal to the summation of the squares of the real part and imaginary parts of the voltage. The formulation objective function is shown in (2).

$$\min \sum_{i \in \Omega_{AMI_V}} \sum_{\phi \in \psi} \left( V_{i,\phi}^{r}{}^2 + V_{i,\phi}^{im}{}^2 - \left| \hat{V}_{i,\phi} \right|^2 \right)^2 \qquad (2)$$

Let $i$ and $j$ be the indices of the sending and receiving buses of a line $(i, j)$. For line flow equations, the mathematical relation between the voltage difference between the two buses $(i, j)$ for each phase $(\phi)$ of a line and the current flow for each phase of a line in an unbalanced three-phase distribution system are expressed in (3)-(4).

$$\sum_{p \in \psi} R_{i,j}^{\phi,p} \left( I_{i,j,p}^{r} + \frac{1}{2} \sum_{k \in \psi} y_{i,j}^{p,k} V_{i,k}^{im} \right) - \sum_{p \in \psi} X_{i,j}^{\phi,p} \left( I_{i,j,p}^{im} - \frac{1}{2} \sum_{k \in \psi} y_{i,j}^{p,k} V_{i,k}^{r} \right) = V_{i,\phi}^{r} - V_{j,\phi}^{r},$$
$$\forall (i,j) \in \Omega_L, \phi \in \psi \qquad (3)$$

$$\sum_{p \in \psi} R_{i,j,}^{\phi,p} \left( I_{i,j,p}^{im} - \frac{1}{2} \sum_{k \in \psi} y_{i,j}^{p,k} V_{i,k}^{r} \right) + \sum_{p \in \psi} X_{i,j}^{\phi,p} \left( I_{i,j,p}^{r} + \frac{1}{2} \sum_{k \in \psi} y_{i,j}^{p,k} V_{i,k}^{im} \right) = V_{i,\phi}^{im} - V_{j,\phi}^{im},$$
$$\forall (i,j) \in \Omega_L, \phi \in \psi \qquad (4)$$

The real and imaginary parts of the current injection constraint are defined using (5) and (6) respectively.

$$I_{i,\phi}^{r,inj} = \sum_{j \in \delta(i)} I_{i,j,\phi}^{r}, \forall i \in \Omega_N, \phi \in \psi \qquad (5)$$

$$I_{i,\phi}^{im,inj} = \sum_{j \in \delta(i)} I_{i,j,\phi}^{im}, \forall i \in \Omega_N, \phi \in \psi \qquad (6)$$

The active power balance constraint is defined in (7). The reactive power balance constraint is defined in (8), with consideration of capacitor output ($Q_{c,\phi}$).

$$\sum_{\forall s \in \Omega_{SB}, \atop s=i} P_{s,\phi}^{G} - \sum_{\forall m \in \Omega_T \atop m=i} P_{m,\phi}^{Tr} - \sum_{\forall l \in \Omega_{D1} \atop l=i} P_{l,\phi}^{D} + \sum_{\forall l \in \Omega_{PV} \atop l=i} P_{l,\phi}^{PV}$$
$$- \sum_{\forall l \in \Omega_{AMI_D} \atop l=i} d_{l,\phi}^{D} = V_{i,\phi}^{r} I_{i,\phi}^{r,inj} + V_{i,\phi}^{im} I_{i,\phi}^{im,inj},$$
$$\forall i \in \Omega_N, \phi \in \psi \qquad (7)$$

$$\sum_{\forall s \in \Omega_{SB}, \atop s=i} Q_{s,\phi}^{G} + \sum_{\forall c \in \Omega_{cap} \atop c=i} Q_{c,\phi}^{C} - \sum_{\forall l \in \Omega_D \atop l=i} Q_{l,\phi}^{D}$$
$$= V_{i,\phi}^{im} I_{i,\phi}^{r,inj} - V_{i,\phi}^{r} I_{i,\phi}^{im,inj}, \forall i \in \Omega_N, \phi \in \psi \qquad (8)$$

where the reactive power output of a connected capacitor is modeled using a constant capacitance model. Therefore, the reactive power is expressed as follows:

$$Q_{c,\phi}^{C} = B_{c,\phi} \left( V_{c,\phi}^{r}{}^2 + V_{c,\phi}^{im}{}^2 \right), \forall c \in \Omega_{cap}, \phi \in \psi \qquad (9)$$

The substation model shown in Fig. 4 is proposed and implemented to capture the unbalanced nature of distribution substations in practical utility feeders (Section III-C). The voltage magnitude limits at the feeder-head are expressed in (10).

$$V_{i,\phi}^{min}{}^2 \leq V_{i,\phi}^{r}{}^2 + V_{i,\phi}^{im}{}^2 \leq V_{i,\phi}^{max}{}^2, \forall i \in \Omega_H, a \neq \phi \in \psi \quad (10)$$

The upper and lower bounds in (10) are established according to the feeder-head available measurements of voltage $\left( \left| \hat{V}_{i,\phi} \right|, i \in \Omega_H \right)$. The feeder modeled in this paper has available hourly measurements of voltage magnitude at the feeder-head for phase $a$ $\left( \left| \hat{V}_{H,a} \right| \right)$, which is considered as a parameter for this phase by the model enhancement algorithm. On the other hand, the voltage magnitude at the feeder-head of phases $b$ and $c$ are calculated individually by the algorithm proposed. A 20% maximum deviation from $\left| \hat{V}_{H,a} \right|$ is set in (10) for phases $b$ and $c$.

The voltage magnitude limits at the substation are expressed in (11). The voltage magnitude of the three phases at the substation are calculated individually by the algorithm proposed with a maximum deviation of 25% from $\left| \hat{V}_{H,a} \right|$.

$$V_{s,\phi}^{min} \leq \left| V_{s,\phi} \right| \leq V_{s,\phi}^{max}, \forall s \in \Omega_{SB}, \phi \in \psi \qquad (11)$$

The real and imaginary part of the voltage at the substation are given by (12) - (13) respectively.

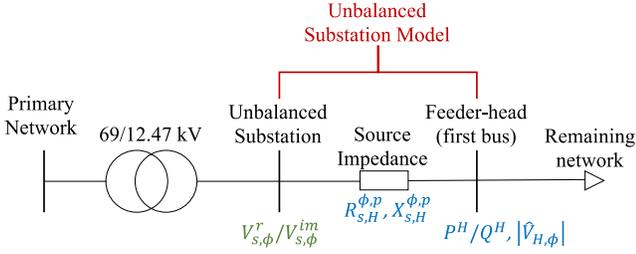

Fig. 4. Substation and feeder-head definition: available measurements are in blue, estimated values are in green. Three independent single phase voltage sources are implemented in the unbalanced substation bus to model an unbalanced circuit.

$$V_{s,\phi}^r = |V_{s,\phi}| \cos(\angle V_{s,\phi}), \forall s \in \Omega_{SB}, \phi \in \psi \qquad (12)$$

$$V_{s,\phi}^{im} = |V_{s,\phi}| \sin(\angle V_{s,\phi}), \forall s \in \Omega_{SB}, \phi \in \psi \qquad (13)$$

where the voltage angles are limited using (14). The angles limits are set to a maximum deviation of $\pm 3°$ from a balanced three-phase source (i.e., $0° \pm 3°$ for phase $a$, $-120° \pm 3°$ for phase $b$ and $120° \pm 3°$ for phase $c$). These limits definitions allow the consideration of an unbalanced source.

$$\angle V_{s,\phi}^{min} \leq \angle V_{s,\phi} \leq \angle V_{s,\phi}^{max}, \forall s \in \Omega_{SB}, \phi \in \psi \qquad (14)$$

The reactive power of each load is limited by its maximum and minimum power factor using (15)-(16).

$$d_{l,\phi}^D \sqrt{\left(\frac{1}{(PF_l^{max})^2} - 1\right)} \leq Q_{l,\phi}^D \leq d_{l,\phi}^D \sqrt{\left(\frac{1}{(PF_l^{min})^2} - 1\right)},$$
$$\forall l \in \Omega_{AMI_D}, \phi \in \psi \qquad (15)$$

$$P_{l,\phi}^D \sqrt{\left(\frac{1}{(PF_l^{max})^2} - 1\right)} \leq Q_{l,\phi}^D \leq P_{l,\phi}^D \sqrt{\left(\frac{1}{(PF_l^{min})^2} - 1\right)},$$
$$\forall l \in \Omega_{D1}, \phi \in \psi \qquad (16)$$

Since DAS measurement of the total three-phase active and reactive power are available at the feeder-head, the summation of power injections at the different phases at the feeder-head is assumed to be equal to the measured value.

$$\sum_{\phi \in \psi} (V_{h,\phi}^r I_{h,\phi}^{r,inj} + V_{h,\phi}^{im} I_{h,\phi}^{im,inj}) = P^H, \forall h \in \Omega_H \qquad (17)$$

$$\sum_{\phi \in \psi} (V_{h,\phi}^{im} I_{h,\phi}^{rinj} - V_{h,\phi}^r I_{h,\phi}^{im,inj}) = Q^H, \forall h \in \Omega_H \qquad (18)$$

### C. Algorithm Output and Simulation Capabilities

The algorithm developed solves a three-phase distribution system power flow problem. The power flow can be solved in standard single snapshot mode and daily variable time-interval mode. The time interval can be any time period. The feeder model developed in this paper is solved for each day using 24-hourly steps. When the power flow is solved, the losses, voltages, flows, and other information are available for the total system and each element.

For each instant in time, the algorithm automatically exports the power flow solution of the system, as well as the loads active and reactive power definition ($P_{l,\phi}^D / Q_{l,\phi}^D, \forall l \in \Omega_D$), distributed generators output ($P_l^{PV}, \forall l \in \Omega_{PV}$), and substation voltage magnitudes and angles ($|V_{s,\phi}|/\angle V_{s,\phi}, \phi \in \psi$) as shown in Fig. 3. These data are then used to complete the OpenDSS model, as explained in the next section.

### D. OpenDSS Optimal Feeder Model

To validate the accuracy of the ACOPF formulation, the resulting distribution model power flow is solved with OpenDSS in a time series solution mode. In this mode, the voltage source and each load and PV generator follow hourly profiles, which are obtained from the enhancement algorithm and are transferred to OpenDSS through shape files. The profiles are created for active and reactive power for each load/PV generator and for the voltage magnitude of each of the three phases at the substation. The loads are modeled using constant $P$ and constant $Q$ to preserve the power flow obtained from the enhancement algorithm. The PVs are modeled as electronically coupled generators using current-limited constant kW based on the AMI measurements, and with reactive power set to zero.

The three-phase voltage sources are modeled by OpenDSS as balanced Thévenin equivalents, that is, a voltage source behind an impedance. However, balanced sources at the distribution network are not accurate to represent the unbalanced voltages that are common in practical utility feeders, since even on a substation with a load tap changing (LTC) transformer, the three voltages on each phase differ with time [25]. Therefore, to model the unbalanced distribution system more accurately, the optimization-based method proposed implements three independent single phase voltage sources at the unbalanced substation bus to model an unbalanced substation source as shown in Fig. 4. These sources are defined in OpenDSS with the actual per unit voltage at which the source (substation) is operating before the source impedance voltage drop. Since the feeder-head measurement of voltage is available only for the magnitude of one phase ($|\hat{V}_{H,a}|$), the feeder-head voltage for the remaining phases, as well as the substation voltages are obtained by the ACOPF formulation proposed. In the enhancement algorithm, the source impedance is represented as a line with no capacitance ($R_{s,H}^{\phi,p}, X_{s,H}^{\phi,p}, \phi, p \in \psi$). Using (10)-(11), the maximum and minimum voltage magnitude at the feeder-head and substation are limited by using $|\hat{V}_{H,\phi}|$ as reference. As a result, three shape files corresponding to the voltage magnitude for each of the three phases at the substation are created for the OpenDSS model. On the other hand, while the enhancement algorithm provides the voltage angle for each phase at each step of the time series simulation, OpenDSS does not allow voltage source angle variation through shapefiles. Therefore, unbalanced angles are defined at each of the three single-phase sources at the substation bus and are kept constant for the entire time series simulation.

Additionally, monitors are set at the feeder-head bus and every load bus in the system to capture the results of voltages and powers at those points. Then, the results of the monitors are compared against the AMI measurements at the loads and DAS measurements at the feeder-head to validate the system. The model validation is discussed in Section IV.

### E. Algorithm- OpenDSS Power Flow Comparison

This section presents the power flow validation between the enhancement algorithm and OpenDSS.

The consumed/produced active and reactive power for each element in the feeder are calculated from the algorithm power flow solution and compared against the corresponding power element losses which are exported from OpenDSS.

The active and reactive power for each element in the feeder are calculated according to the type of the element. For all the lines ($\Omega_L$) in the feeder, where $i$ and $j$ are the indices of the sending and receiving buses, the active and reactive power consumption (inductive lines), and reactive power production (capacitive underground lines) are calculated as the power difference between the sending and receiving ends.

Similarly, for all the transformers ($\Omega_T$) in the feeder, the active and reactive power consumption are calculated as the power difference between the sending and receiving ends plus the no-load loss ($P_m^{Tr}, \forall\, m \in \Omega_T$), which represents a resistive branch in parallel with the magnetizing inductance. For all the capacitors ($\Omega_{cap}$), (9) is used to calculate the reactive power injections.

The load active power ($d_l^p, \forall l \in \Omega_{AMI_D}$; $P_l^D, \forall l \in \Omega_{D1}$), and reactive power ($Q_l^D, \forall l \in \Omega_D$) are directly calculated in the algorithm using (7)-(8). The PVs' active power production ($P_l^{PV}, \forall l \in \Omega_{PV}$) are inputs from the AMI data and are assumed to operate at unity power factor.

The source active and reactive powers are calculated by using the receiving end (feeder-head end) of the line that connects the substation with the feeder-head, that is,

$$P_{s,\phi}^G = V_{H,\phi}^r I_{s,H,\phi}^r + V_{H,\phi}^{im} I_{s,H,\phi}^{im}, \phi \in \psi \quad (18)$$

$$Q_{s,\phi}^G = V_{H,\phi}^{im} I_{s,H,\phi}^r - V_{H,\phi}^r V_{H,\phi}^{im}, \phi \in \psi \quad (19)$$

TABLE I shows the comparison between results of the enhancement algorithm and the OpenDSS model based on it for the total consumed/produced power for different types of elements during a single snapshot (historical feeder load peak). The corresponding comparisons match under 0.13% error (except for the transformer active power comparison, which has an error of 2.95%, equivalent to 3 kW). For the same snapshot, Fig. 5 shows a per phase comparison of the voltage magnitude of all the buses ($\Omega_N$) between the enhancement algorithm and OpenDSS solution. Fig. 6 shows the voltage errors between OpenDSS and the enhancement algorithm for time-series analysis, for different hours of a single day (historical feeder load peak day). The results show that the enhancement algorithm models the power flow constraints correctly and that the results obtained from the enhancement algorithm match closely with the electrical model and assumptions employed in a state-of-the-art distribution system power flow solver such as OpenDSS.

### IV. Model Validation: Utility Feeder Results

This section presents the results for an actual 12.47 kV, 9 km-long Arizona utility feeder that serves residential customers. Fig. 7 shows the circuit diagram of the feeder with all its elements. Peak net load on the feeder was 7.35 MW on 07/15/2019. The feeder has one of the highest PV penetrations among the utility's operational feeders with 3.8 MW of residential rooftop PV installed. This represents a penetration level of more than 200% (3.8 MW/1.6 MW) as compared to the feeder total gross load during peak solar PV production hours. The OpenDSS model for this feeder has an unbalanced 69/12.47 kV source representing the substation, 7864 buses, 1790 primary sections, 5782 secondary sections, 371 distribution transformers, four capacitor banks of 1.2 MVAr

TABLE I
Comparison of active and reactive power from Power Flow Solution

| Component | Proposed Algorithm | | OpenDSS | | % Error | |
|---|---|---|---|---|---|---|
| | P (MW) | Q (MVar) | P (MW) | Q (MVar) | P | Q |
| Source | -5.978 | -0.522 | -5.979 | -0.523 | 0.01 | 0.13 |
| Lines | 0.083 | -0.423 | 0.083 | -0.423 | 0.17 | 0.06 |
| Capacitors | 0.000 | -1.261 | 0.000 | -1.260 | 0.00 | 0.01 |
| Loads | 7.725 | 2.166 | 7.728 | 2.167 | 0.03 | 0.02 |
| Transformers | 0.073 | 0.040 | 0.070 | 0.040 | 2.95 | 0.01 |
| PVs | -1.903 | 0.000 | -1.903 | 0.000 | 0.00 | 0.00 |

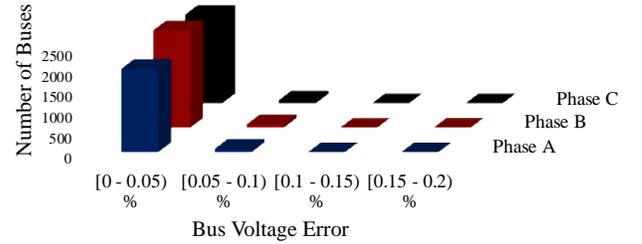

Fig. 5. Bus voltage magnitude error per phase between enhancement algorithm power flow and OpenDSS solution.

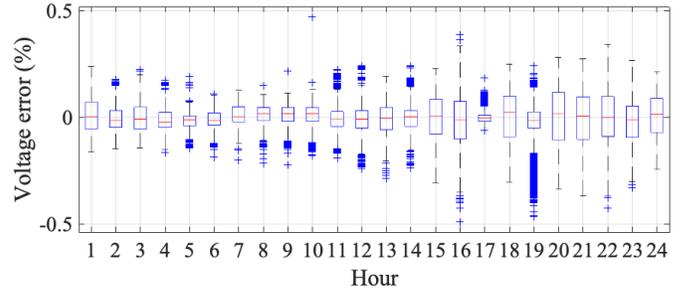

Fig. 6. Voltage errors between OpenDSS and the optimization-based algorithm for different hours of a single day.

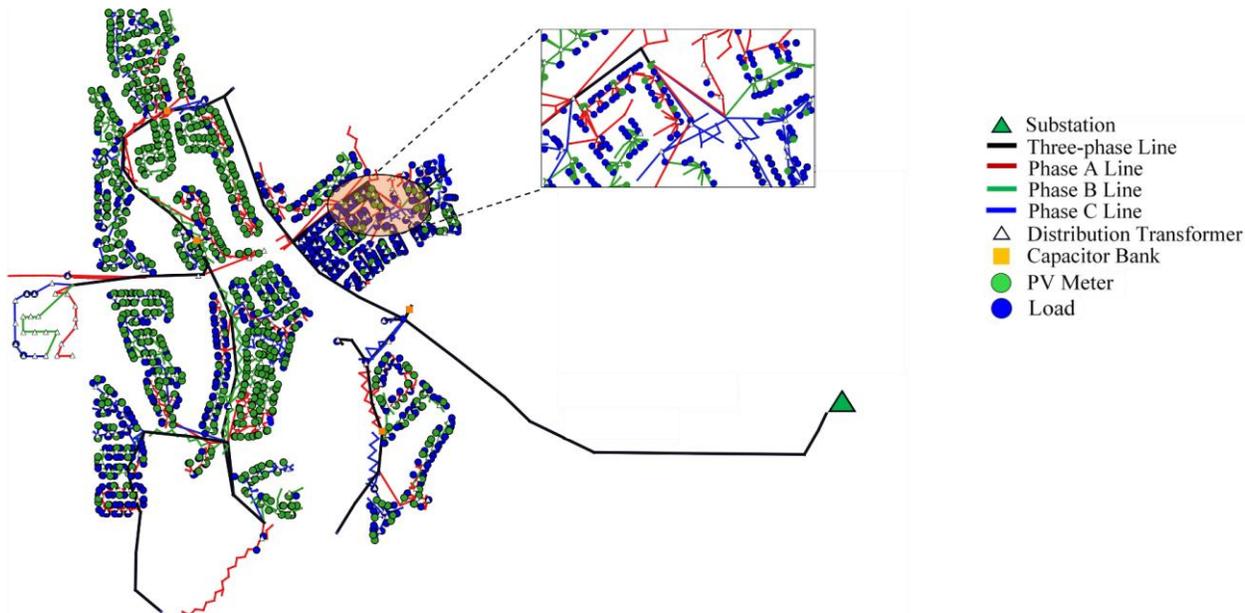

Fig. 7. Arizona utility feeder circuit diagram with all the elements.

rating each, 1737 loads, and 766 PV units.

To validate the feeder model, two days corresponding to the actual historical feeder load peak on 07/15/2019 (high load and relatively low PV), and the maximum generation condition on 03/15/2019 (high PV) were chosen for the analysis.

Following the procedure described in the previous sections, the feeder model is tuned and an updated OpenDSS model is created as a result. Using this updated OpenDSS model, a time series power flow yielded a good match with the measured values. The OpenDSS feeder-head active and reactive powers are compared with the corresponding feeder-head measurements for both days in Fig. 8 and Fig. 9. For the historical feeder load peak day, the feeder-head active and reactive powers have root mean square (RMS) errors over a day of 0.104% and 0.234% respectively. For the day with the maximum generation, the feeder-head active and reactive powers have a root mean square (RMS) errors over a day of 1.897% and 0.0493% respectively. Note that the reactive power is wholly calculated by the optimization-based technique as there are no reactive power measurements available at any point along the feeder aside from the feeder-head values. The active power is varied for a subset of the loads ($\Omega_{D1}$) by the enhancement algorithm, therefore, a small error at the feeder-head active power is an indication of the model successfully being tuned to represent the measurements and, by proxy, the actual feeder status.

As a further validation, the voltages at the premises along the feeder where AMI measurements are available are compared against these measurements. This comparison between the AMI and model voltages for three representative meters at different locations along the feeder for the historical feeder peak load day and maximum generation condition day are shown in Fig. 10 and Fig. 11 respectively. The distances from the substation of the locations corresponding to the plot are shown above each plot. The RMS error over a day is also calculated for all the meters along the feeder. Fig. 12 shows the RMS error calculated over a day for both days in a box-and-whisker plot. As observed, the average RMS error along the feeder length is around 0.4 %, for the historical feeder load peak day and 0.2% for the maximum generation condition day, which shows that the proposed method achieves a very good match when compared against field measurements.

Fig. 13 shows the AMI measurements received from the utility compared against the corresponding voltage profile obtained from the enhancement algorithm. The method proposed provides a complete power flow solution for both primary and secondary side by using sparse measurements at secondary level along the feeder, extending the observability and planning capabilities of the feeder under study. Due to the accuracy of the model obtained, these results are currently being used as input data for other studies such as topology processor, distribution system state estimation (DSSE), and optimal energy management of DERs.

## V. DETAILED FEEDER CHARACTERISTICS

This section presents some feeder characteristics derived from the detailed feeder model developed.

Fig. 14 shows the gross load, net-load, and PV production for both the historical feeder load peak day and maximum generation condition day. For the maximum generation condition day, it is seen that the PV production significantly impacts the net load of the system. In this case due to the significant penetration of solar PV, the net load is negative between 10 AM-3 PM. Due to this behavior, the feeder experiences large overvoltage during this time. The maximum solar PV generation occurs around 1 PM. The voltage profile for this time is shown on the right-hand side in Fig. 15, and it indeed shows a trend of increasing magnitude moving away from the substation. This trend is noteworthy since it goes counter to the traditional assumptions of a distribution system with little or no active generation. Also note that the unbalance between the phases in the feeder is successfully captured by the optimization-based technique proposed.

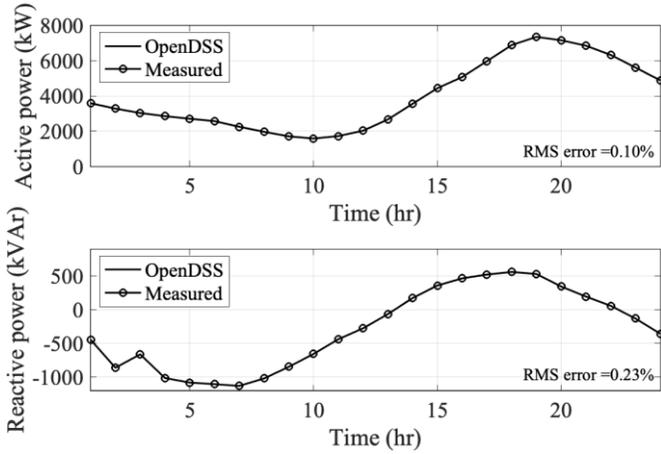

Fig. 8. Active and reactive power feeder-head comparison between OpenDSS detailed model and DAS measurements for the historical feeder load peak day.

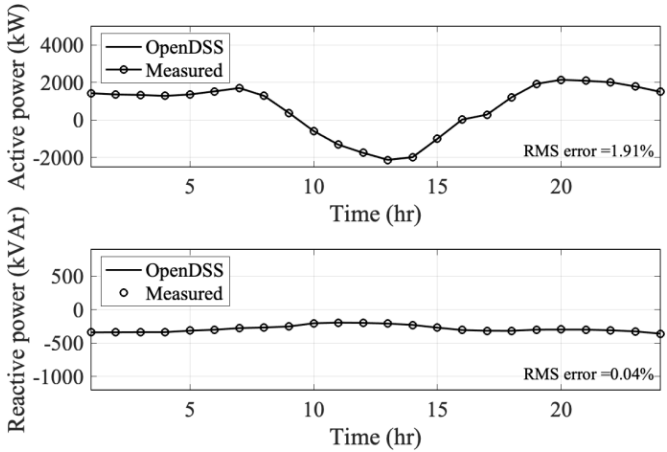

Fig. 9. Active and reactive power feeder-head comparison between OpenDSS detailed model and DAS measurements for the maximum generation condition day.

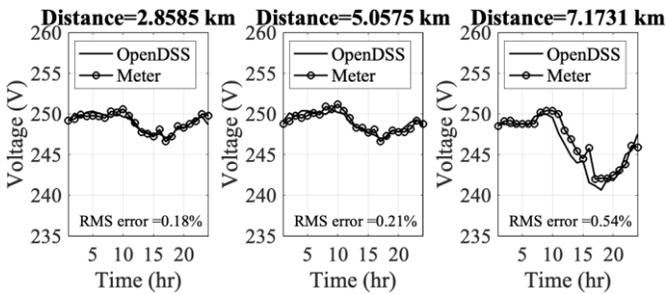

Fig. 10. Voltage comparison results between OpenDSS and AMI data for some premises along the feeder for the historical feeder load peak day.

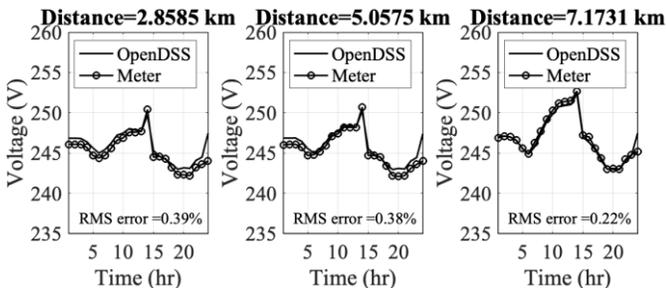

Fig. 11. Voltage comparison results between OpenDSS and AMI data for some premises along the feeder for the maximum generation condition day.

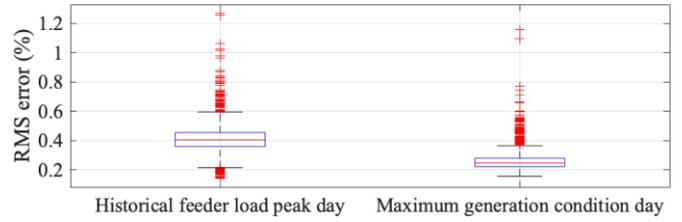

Fig. 12. RMS error along the feeder over a day

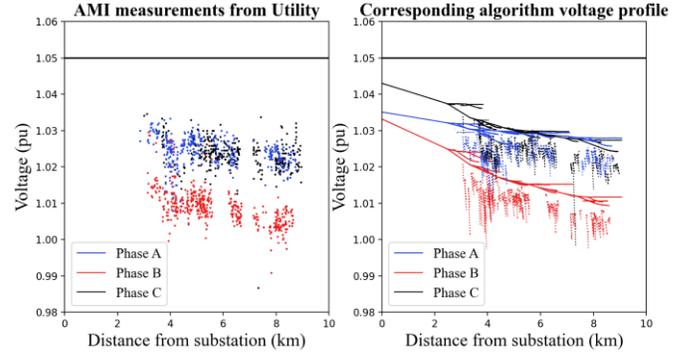

Fig. 13. AMI measurements from the utility compared against the corresponding voltage profile obtained from the enhancement algorithm.

As shown in Fig. 14, the historical feeder load peak day has high load, which is why there is no reverse flow at the feeder-head – there is still a significant reduction in the net load due to solar PV production. In Fig. 15 the voltage profile of the feeder corresponding to an evening condition (high load) for the historical feeder load peak day is plotted on the left-hand side. This voltage profile shows a decreasing voltage trend as we go away from the substation as is traditionally expected for distribution systems. The contrast in the feeder voltage profiles between the two snapshots presented in Fig. 15 as well as the gross and net loads presented in Fig. 14 highlight the fact that a distribution system with a high solar PV penetration can exhibit a wide range of behaviors, hence accurately modeling the distribution system is important for any studies involving it.

Fig. 16 shows the power factor for some premises along the feeder for the historical feeder load peak day. The enhancement algorithm estimates the power factor for each load independently on the location or time.

## VI. CONCLUSION

To accurately represent the distribution systems in studies, constructing a detailed model which corresponds to the actual feeder(s) is essential. This paper describes a novel procedure to enhance the model of a real distribution system feeder using AMI and DAS data. This novel enhancement algorithm formulates an ACOPF based on IV formulation for matching the voltages at various nodes to AMI measurements, ensuring that the tuned model closely reflects the real-life status of the feeder for a snapshot. The model is validated using time-series analysis in OpenDSS. Simulation results for voltages have an average RMS error along the feeder under 0.5%, and all RMS errors are under 1.4%, when compared to the field measurements providing confidence in the developed method. In the future work, this modeling procedure can be used to create extended databases for further analysis and study of the distribution network when no AMI data is available to construct the profiles, and guide distribution system future extension.

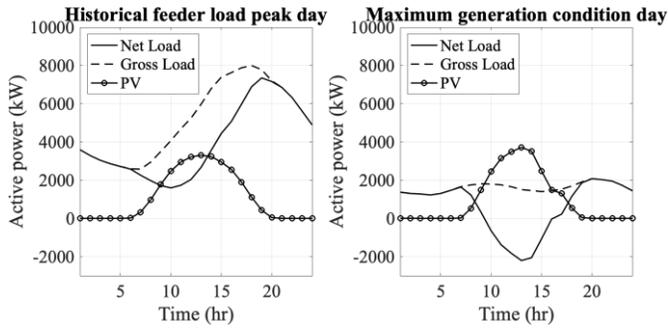

Fig. 14. Gross load, net load, and PV production behavior.

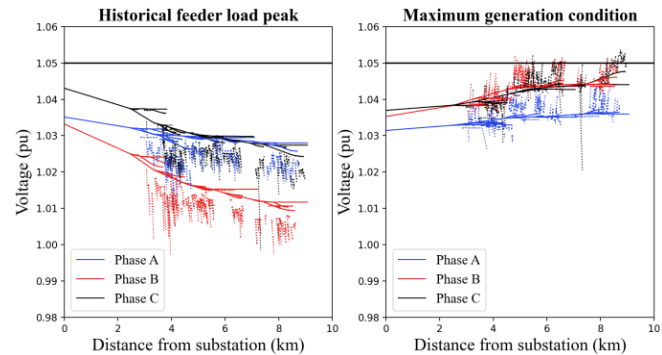

Fig. 15. Voltage profile comparison between the historical feeder load peak and the maximum generation condition.

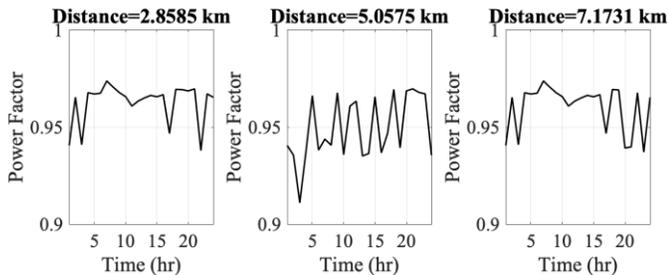

Fig. 16. Power factor for some premises along the feeder for the historical feeder load peak day.